\documentclass[
aip,
apl,
graphicx,
amsmath,amssymb,
reprint,
]{revtex4-1}
\usepackage{graphicx}
\usepackage{float}

\draft 

\begin{document}


\title{Demonstration of a superconducting nanowire microwave switch} 



\author{A. Wagner}
\email[]{andrew.wagner@raytheon.com}
\affiliation{Raytheon BBN Technologies,Cambridge,MA 02138, USA}

\author{L. Ranzani}
\affiliation{Raytheon BBN Technologies,Cambridge,MA 02138, USA}

\author{G. Ribeill}
\affiliation{Raytheon BBN Technologies,Cambridge,MA 02138, USA}

\author{T. A. Ohki}
\affiliation{Raytheon BBN Technologies,Cambridge,MA 02138, USA}


\date{\today}

\begin{abstract}
The functionality of a nanowire integrated into a superconducting transmission line acting as a single pole single throw switch is demonstrated. The switch has an instantaneous bandwidth from 2 to 8 GHz with more than 10 dB of isolation between the open and closed states. The switch consumes no power in the closed state and $\approx 15~\rm{nW}$ in the open state. The rise and fall response time between open and closed states is approximately $370~\rm{ps}$.
\end{abstract}

\pacs{\#}

\maketitle 

Quantum computing architectures employing as many as 72 qubits have recently been demonstrated. In order to scale such
architectures further, miniaturized circulators, isolators, and switching networks that have low power consumption while
operating at cryogenic temperatures will be necessary~\cite{Hornibrook:2015kh}. Several groups have demonstrated
microwave switches and phase shifters based on Josephson junctions~\cite{Naaman:2016fj,Naaman:2017gh,Chapman:2016fy,Kokkoniemi:2017er,Brummer:2011iu}, and novel semiconductor devices~\cite{Hornibrook:2015kh,Borodulin:2018cq}. A cryogenic switch operating at DC, based on a cryotron has also been reported~\cite{Lowell:2016ks}. In this letter, we
demonstrate a single pole single throw (SPST) switch fabricated from a nanowire integrated into a superconducting
transmission line as an alternative low power, cryogenic microwave switch. Such nanowires are commonly used in the
fabrication of superconducting single photon detectors~\cite{Kadin:1996hy,Goltsman:2001,Engel:2004ka,Semenov:2001bv}, and a three
terminal variant of the device has been shown to operate as a transistor for digital logic
applications~\cite{McCaughan:2014it}. We present two variants of the switch, one with a single nanowire and a
second with two nanowires operated in tandem to provide improved isolation between the open and closed states of the
switch. The single nanowire device is a small, $w=80~\rm{nm}$ wide nanowire integrated into a superconducting
transmission line with two on-chip inductors fabricated in a single metal layer. Surface mount capacitors
are soldered off chip onto the transmission line feeding the device, creating a bias tee. Modulation of the switch is
achieved by applying a low frequency signal to the inductive ports of the bias tee of sufficient power to exceed the
critical current of the nanowire. The switch is in the closed state when the nanowire is superconducting and forms a
lossless transmission line. Similarly the switch is open when the nanowire has been driven into the normal state.
The switches are fabricated on a single layer of $d=8~\rm{nm}$ thick NbN with a sheet resistance of
$\approx356~\Omega/$square. A wire only $3$~squares long fabricated from this material will produce an impedance of
$\approx1~\rm{k}\Omega$ in the resistive state, enough to obtain more than $20$~dB of RF isolation. In the tandem 
nanowire design a second nanowire increases isolation by shorting one RF port to ground while the first nanowire is
simultaneously in the normal state. 

\begin{figure}[H]
\includegraphics[keepaspectratio=true,width=\columnwidth]{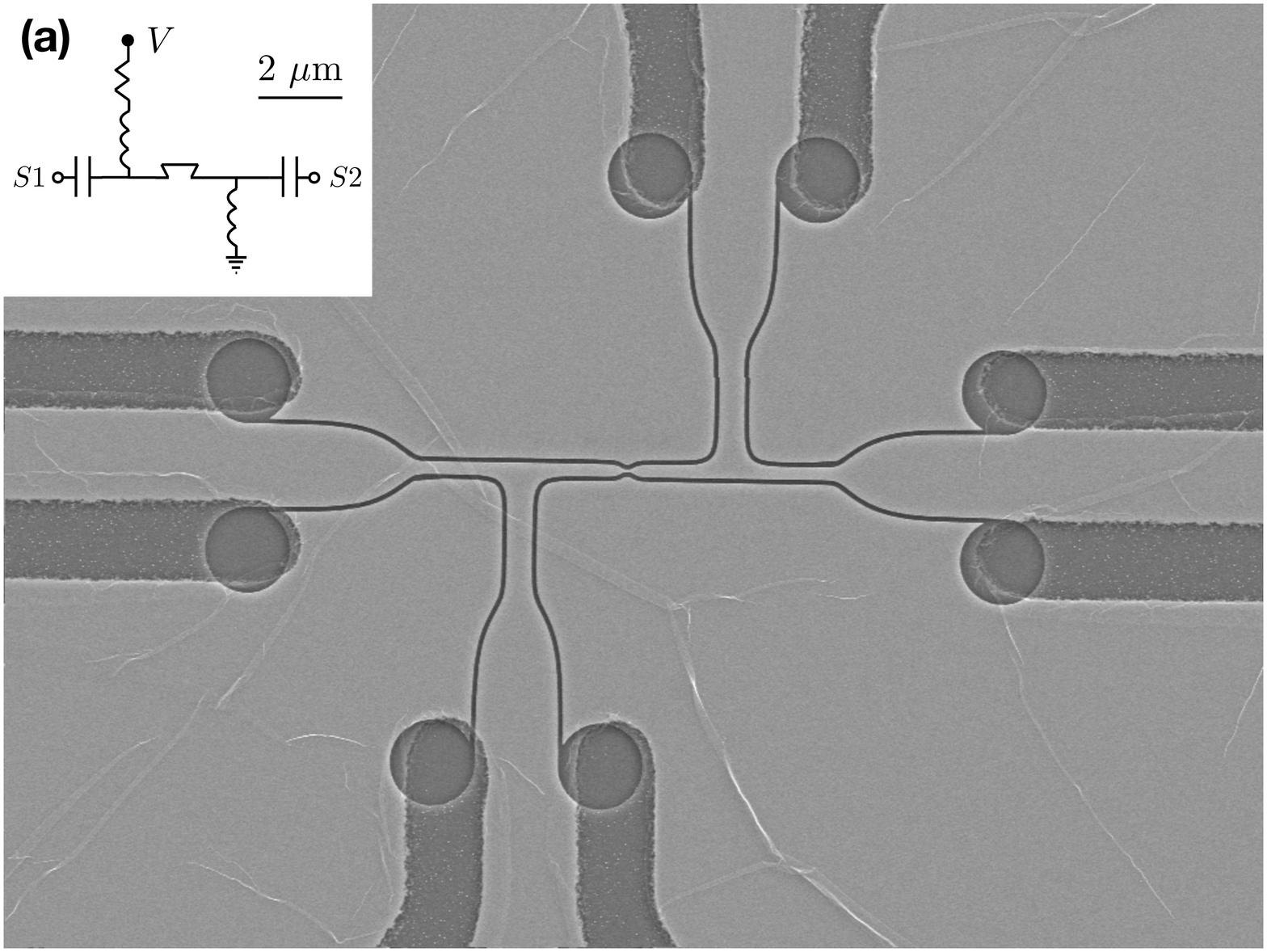}

\vspace{0.1cm}

\includegraphics[keepaspectratio=true,width=\columnwidth]{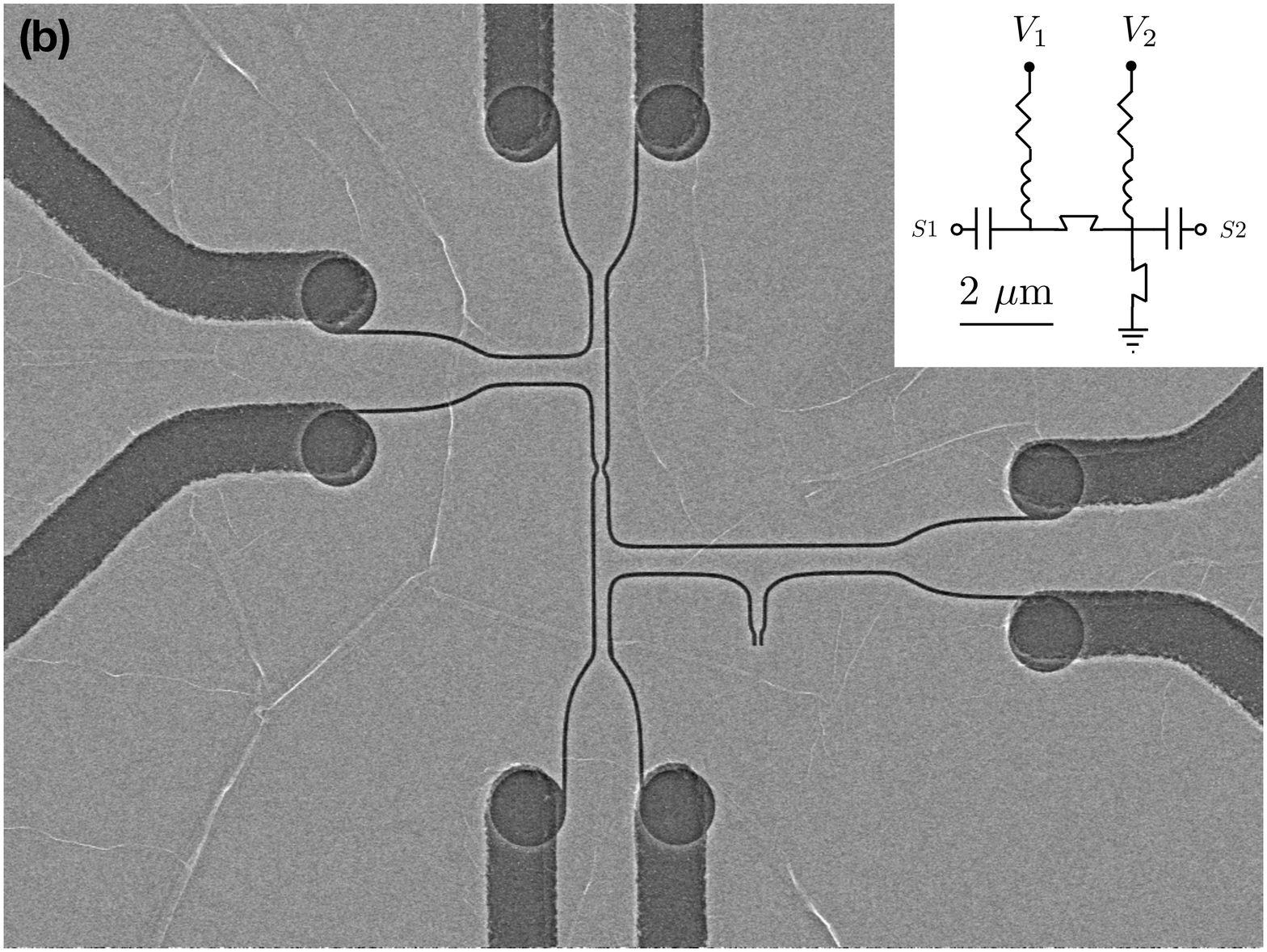}
\caption{\label{fig:circuit} SEM micrograph images and circuit diagrams of the single nanowire switch (a) and tandem nanowire switch (b). The single nanowire switch is actuated by a voltage signal $V$ that drives current through a resistor connected to the inductors of the bias tee. The tandem nanowire switch is actuated with a differential signal across two inductors. In the closed state of the switch $V_2=V_1$ and the control current flows through the shunt nanowire, which switches to the normal, high resistance, state. In the open state of the switch, $V_2-V_1>0$ and the control current flows through the series nanowire. We achieve high RF isolation when the series nanowire is in a high resistance state, while the shunt nanowire is superconducting.}
\end{figure}

\noindent The equivalent circuit of the two switch designs is illustrated by inset circuit diagrams on corresponding SEM images of the fabricated devices in Fig.~\ref{fig:circuit}.

\begin{figure}[H]
\includegraphics[width=\columnwidth]{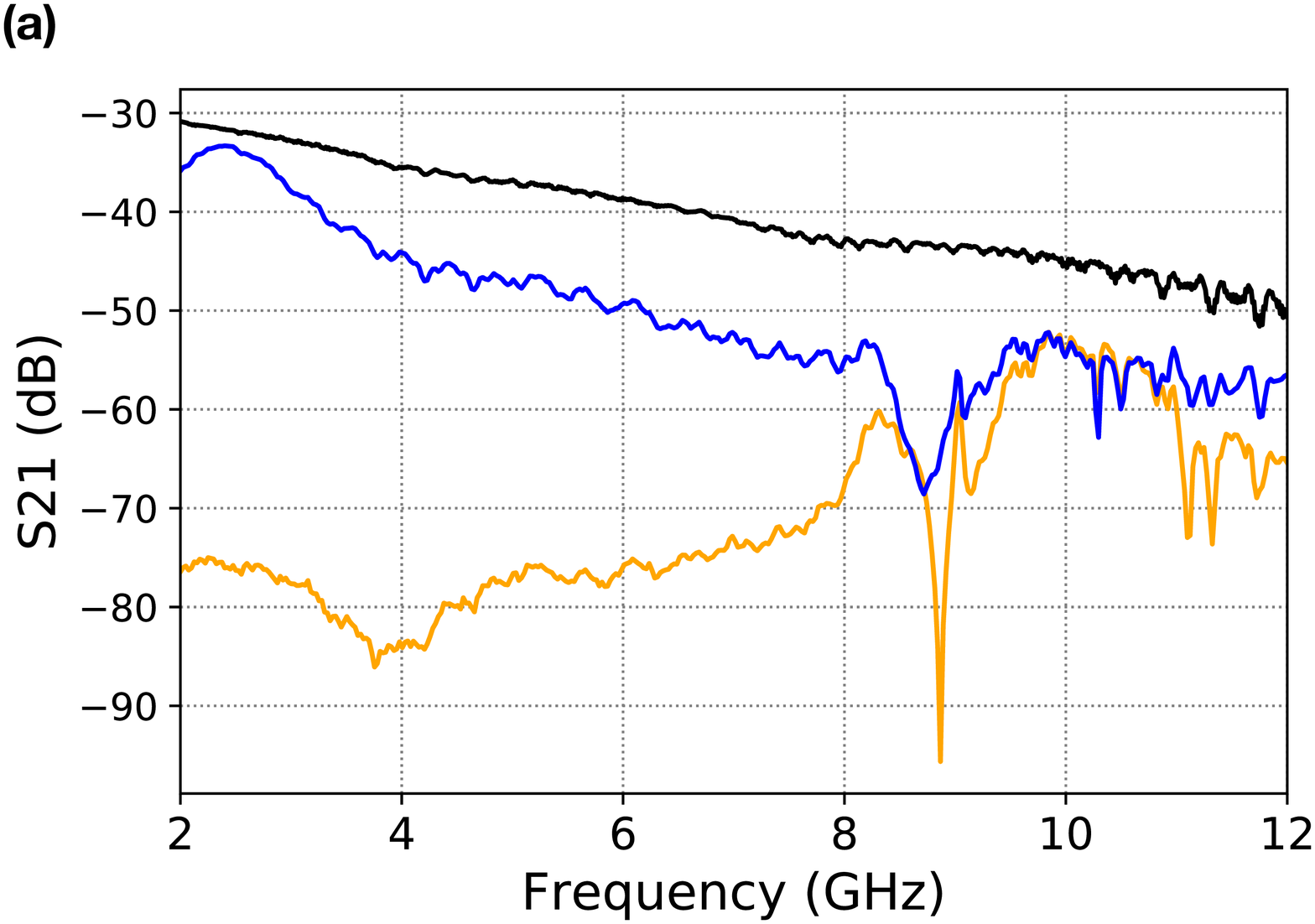}

\vspace{0.1cm}

\includegraphics[width=\columnwidth]{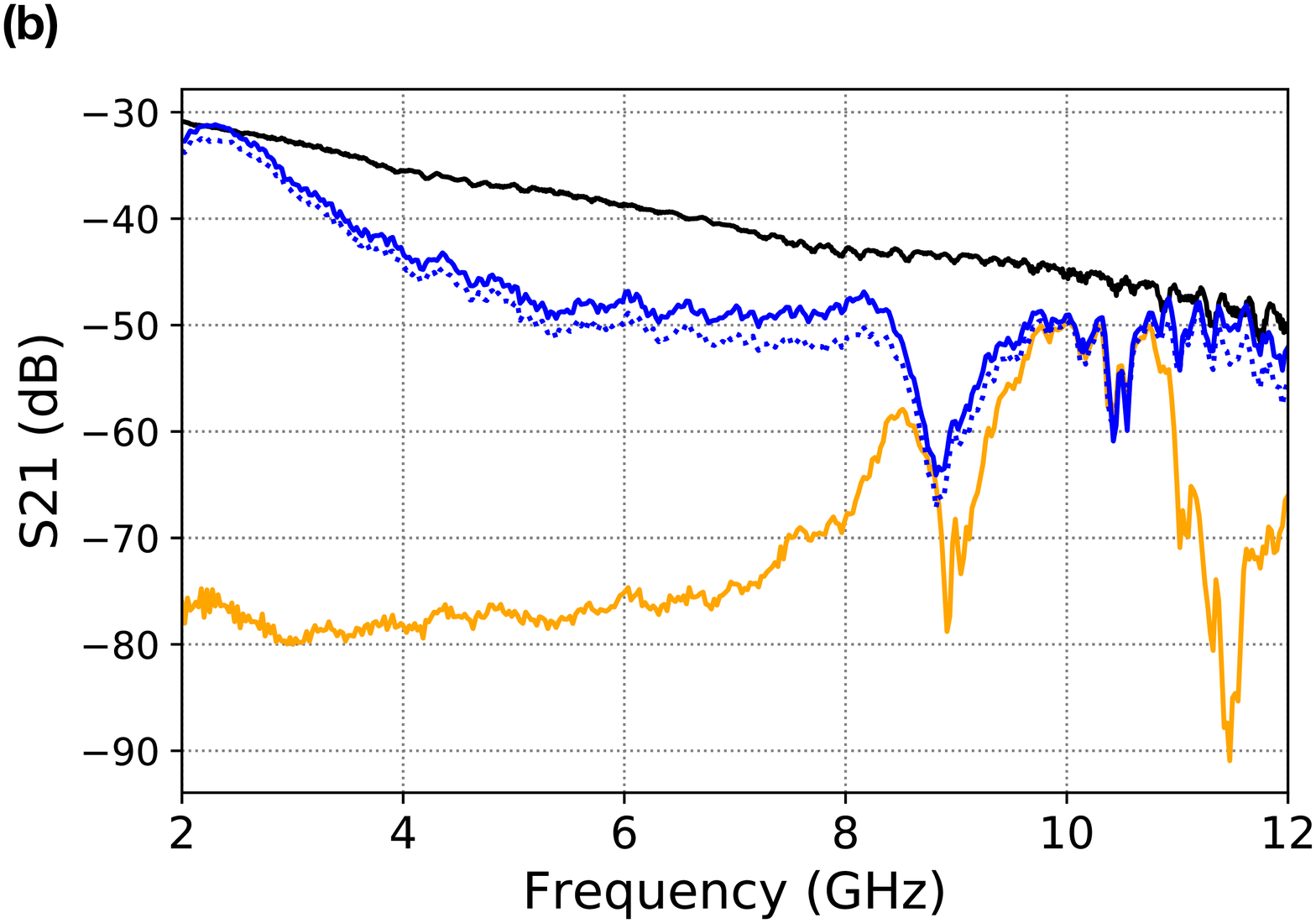}
\caption{\label{fig:transmission} Transmission (S21) of the single nanowire (a) and tandem nanowire (b) in the closed (blue) and open (orange) states. The transmission of the tandem nanowire when disconnected from DC bias current is shown in dashed blue. The transmission of a length of superconducting coaxial cable (black) using the same lines, attenuators and amplifiers as the switches is shown for comparison and was measured in a separate cool-down.}
\end{figure}

In order to engineer the transmission line and integrated bias tee we must account for the kinetic inductance per unit square of 
the thin NbN film, given in the zero temperature ($T=0$), thin film limit by Eq.(\ref{eqn:kineticind})~\cite{Zmuidzinas:2012kh}. 

\begin{equation}
\label{eqn:kineticind}
 L_k=\frac{\hbar}{\pi\Delta_0}\frac{\rho_n}{d}
\end{equation}

\noindent Here $\Delta_0=1.76~k_B T_c$ is the superconducting energy gap, $\rho_n$ is the normal state 
resistivity and $d$ is the film thickness. The high kinetic inductance of superconducing films of the type presented here dominates the inductive contribution to the circuit impedance in the microwave regime. Detailed discussions of the temperature and current dependence of the kinetic inductance are presented in Refs.~\onlinecite{Clem:2012df,Zmuidzinas:2012kh}. We measure a transition temperature $T_c = 8.9~\rm{K}$ and normal state resistivity $\rho_n = 284.8~\mu\Omega\cdot\rm{cm}$ of the NbN films which we pattern into the nanowires, transmission lines and inductors.

We use finite element modeling to find that a coplanar transmission line with a $2~\mu\rm{m}$ gap and $66~\mu\rm{m}$ 
width has $50-64~\Omega$ characteristic impedance for a line kinetic inductance in the range of $25-50$~pH per square, dimensions 
which present no fabrication challenges. The propagation velocity in a high kinetic inductance transmission line is suppressed by the additional inductance and we estimate it to be $4.65\times10^7~\rm{m/s}$ in the devices presented here. Despite the reduced wavelength of propagation in the line, the $80~\rm{nm}$ wide, $160~\rm{nm}$ long nanowire switching elements are well approximated as a small lumped-element inductance in the superconducting state and a large lumped-element resistance in the normal state. The switch devices consist of $400~\mu\rm{m}$ long transmission lines leading to the nanowire switching element. The bias tee consists of two $50$~nH inductors patterned from $280~\mu\rm{m}$ long $40~\mu\rm{m}$ wide meanders of NbN into the same metal layer as the switch, in addition to two $10$~pF off chip capacitors connected by wire bonds to the transmission lines. The fabrication of the switches consists of two lithographic patterning steps and etches. The transmission lines and bias tee inductors are patterned in a photolithographic step while the nanowire switch elements are patterned using ebeam lithography. An overlap between these two patterning steps is necessary for alignment and is visible in Fig.~\ref{fig:circuit}.

\begin{figure}[H]
\includegraphics[width=\columnwidth]{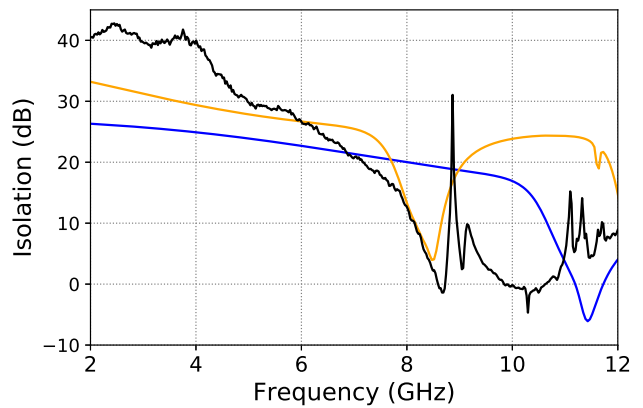}
\caption{\label{fig:isolation} The ratio of S21 between open and closed states of the single nanowire switch (isolation) is shown for the measured device (black) and simulation assuming $50$~pH/sq (blue) and $100$~pH/sq (orange) sheet inductance. Finite element software is used to simulate the isolation of the device.}
\end{figure}

\begin{figure}[H]
\includegraphics[width=\columnwidth]{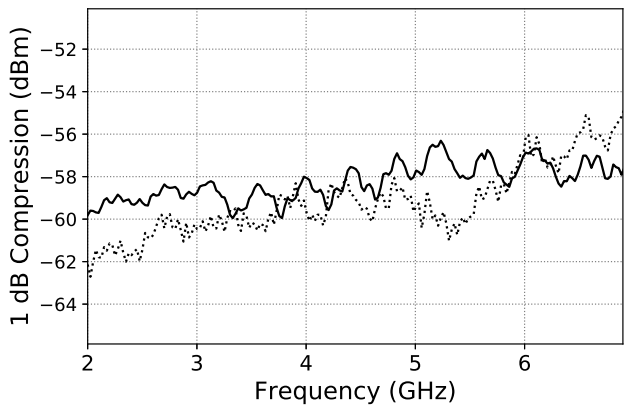}
\caption{\label{fig:compression} $1~\rm{dB}$ compression point as a function of frequency in the closed state for the single nanowire (solid) and tandem nanowire (dashed) switch designs.}
\end{figure}

We use a network analyzer to measure the broadband microwave response of the switches in the open and closed states in Fig.~\ref{fig:transmission}. The switches are opened by applying a DC voltage bias through 33 dB of attenuation, sufficient to exceed the critical current, $I_c \approx 6~\mu A$ of the nanowires. The isolation expected from a finite element model of a $1~\rm{k}\Omega$ resistor embedded in a transmission line is compared to the measured isolation in Fig.~\ref{fig:isolation}. This model is a mesh of the pattern used to define the switch and accounts for any parasitic effects of the switch design. We find that the isolation is consistent with a film of $\approx 100$~pH per square, approximately twice the value we expected from the measured sheet resistance and transition temperature of the unpatterned film. We attribute this increase in kinetic inductance to the fabrication process which exposes the thin NbN film to two lithographic patterning and etch steps as well as an oxygen plasma cleaning step. Notably, the resulting mismatch in line impedance from the design value results in a high reflection coefficient which accounts for the high insertion loss measured in Fig.~\ref{fig:transmission}. We also measure the $1~\rm{dB}$ compression point of the output power for each switch as a function of frequency in the closed state in Fig.~\ref{fig:compression}. We note that in the open state the switch is highly reflective so the transmission in the open state degrades at a much higher power than that in the closed state. 

\begin{figure}[h]
\includegraphics[width=\columnwidth]{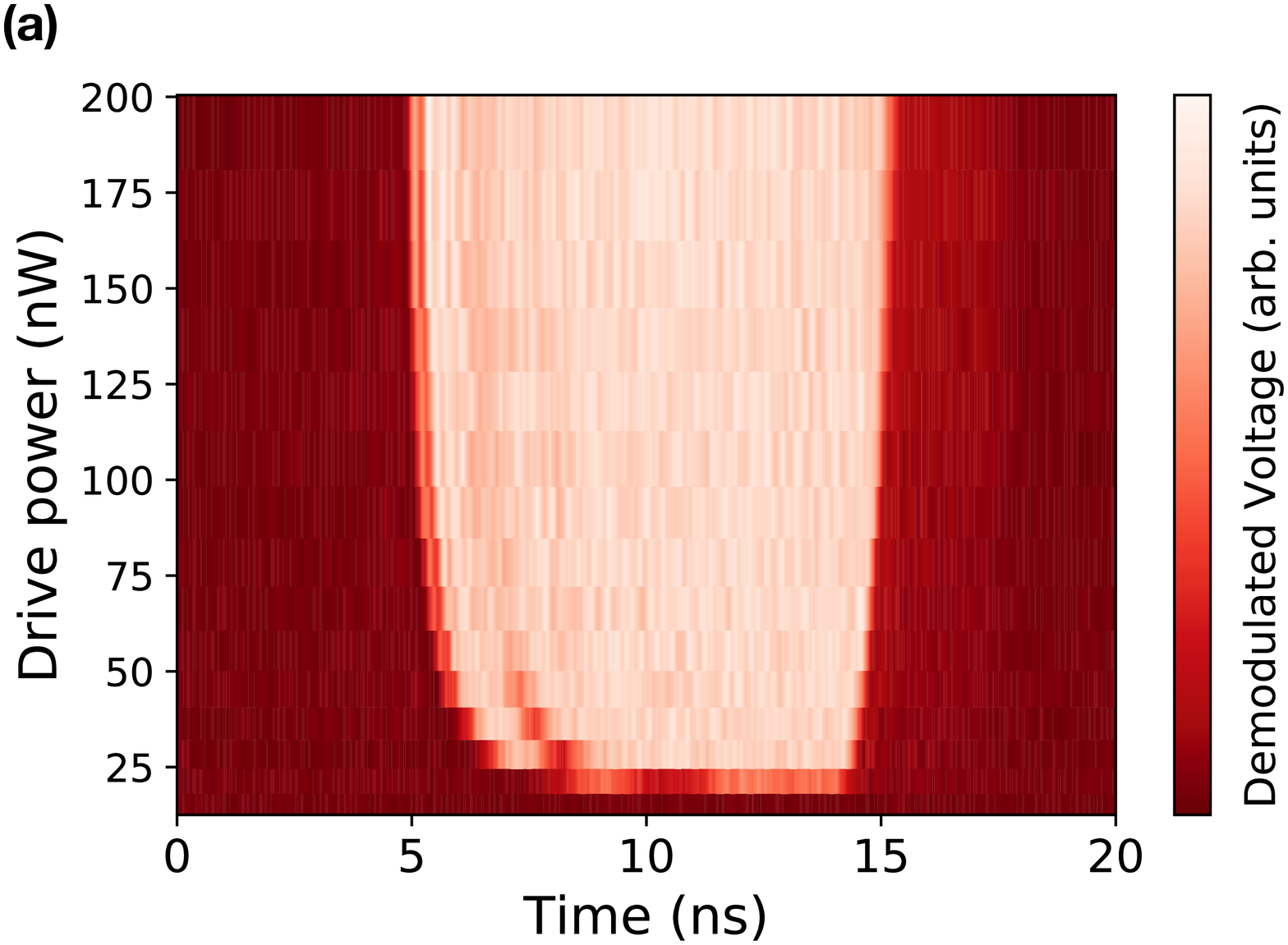}

\vspace{0.1cm}

\includegraphics[width=\columnwidth]{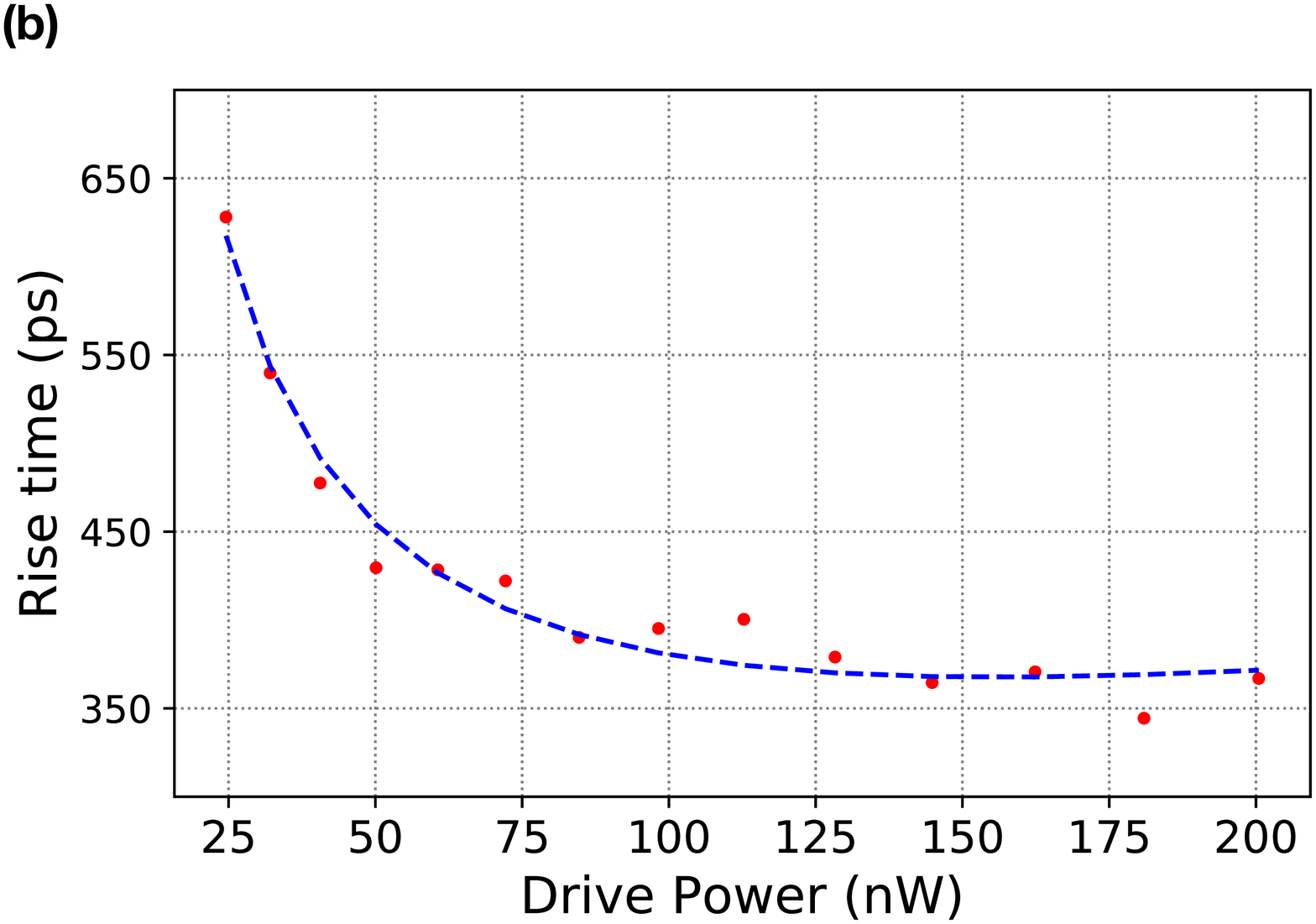}
\caption{\label{fig:homodyne} Response of the single nanowire switch over a complete $50~\rm{MHz}$ drive cycle (a) and rise 
time of the response (b) as a function of driven modulation power. The false color indicates the switch is completely closed in red 
and completely open in white. The rise time of the demodulated signal is measured from the first rising edge of the switching pulse and is shown in red dots. The result of the fit to the electro-thermal model is shown in dashed blue.}
\end{figure}

Finally the switching time and response of the single nanowire switch is measured as function of modulation power is shown in Fig.~\ref{fig:homodyne}a, using a homodyne measurement technique similar to that in Ref.~\onlinecite{Chapman:2016fy}. A $6~\rm{GHz}$ signal is split into two components. The first component is sent through the switch and then combined with the second component into a mixer with a $10~\rm{GHz}$ IF bandwidth. At the same time, a $50~\rm{MHz}$ drive tone is used to modulate the switch between the open and closed states. The demodulated output of the mixer is low pass filtered below $5~\rm{GHz}$ and measured on a fast $20~\rm{GS/s}$ oscilloscope. 

We fit the measured switching time of the nanowires to an analytic electro-thermal model for the rise time of the switch isolation that includes the response of the inductors in the bias tee and the propagation speed of the hot spot in the nanowire. The velocity of the hot spot propagation is given in Eq.(\ref{eqn:hot spot})~\cite{Gurevich:1987zz}. 

\begin{equation}
\label{eqn:hot spot}
v=v_0\frac{\psi i^2 -2}{\sqrt{\psi i^2-1}},~i=\frac{I}{I_c}
\end{equation}

\noindent Here the dimensionless parameter $\psi$ is the ratio of heat dissipated in the resistive hot spot in the wire to the 
heat transported away by contact with the substrate, and $I$ is the drive current. In terms of the superconducting film parameters $\psi$ is given in Eq.(\ref{eqn:steakly}) by 

\begin{equation}
\label{eqn:steakly}
\psi = \frac{\rho_n j_c^2 d}{\alpha(T_h-T_s)}
\end{equation}

\noindent for a critical current density $j_c$, hot spot temperature $T_h$, substrate temperature $T_s$, and thermal contact conductivity $\alpha~(\rm{W/Km^2})$. The normal state resistivity $\rho_n$, and film thickness $d$ are defined in Eq.(\ref{eqn:kineticind}). The characteristic hot spot velocity $v_0~(\rm{m/s})$ is given in Eq.(\ref{eqn:velocity}) by

\begin{equation}
\label{eqn:velocity}
v_0 = \frac{1}{c}\sqrt{\frac{\kappa\alpha}{d}}
\end{equation}

\noindent for thermal conductivity $\kappa = 2.45\times 10^{-8}~\frac{T_h}{\rho_n}~\rm{W/mK}$ found from the Wiedemann-Franz law. The heat capacity, $c~(\rm{J/Km^3})$ is modeled as the temperature dependent sum of contributions for electrons and phonons in the film~\cite{Yang:2007cu}. The response of the nanowire can be modeled by considering a voltage produced from a $Z_0=50~\Omega$ source and delivered to a $50~\Omega$ terminator where the nanowire is between the source and terminator. The voltage produced by such a source, delivering a power $P$, is $V=\sqrt{8 Z_0 P}$. We now consider the total resistance of the circuit as a function of time and drive current,

\begin{equation}
\label{eqn:resistance_1}
\frac{V}{I} = 2Z_0 + \frac{\rho_n}{wd}\int{2v~dt}
\end{equation}

\noindent which can be recast as the differential equation, 

\begin{equation}
\label{eqn:resistance_2}
\frac{V}{I^2} \frac{dI}{dt} = -\frac{2\rho_n}{wd}v~,
\end{equation}

\noindent where $w$ is the width of the nanowire. The factor of $2$ in Eqs.(\ref{eqn:resistance_1},\ref{eqn:resistance_2}) is due to the fact that the hot spot propogates symmetrically in each direction. The time required for the hot spot to expand to its final size can be found by integrating Eq.(\ref{eqn:resistance_2}) from the initial current flowing in the nanowire $I_0=V/2Z_0$ to the steady state current $I_s=I_c\sqrt{2/\psi}$. The steady state current is found by solving Eq.(\ref{eqn:hot spot}) when the hot spot velocity, $v=0$. The rise time of the nanowire switch is the sum of the time required for the hot spot to expand and the time required for the bias tee inductors to charge to the steady state current, 

\begin{equation}
\label{eqn:inductor}
1-e^{-\frac{t}{\tau}} = \frac{I_s}{I_0}
\end{equation}

\noindent where $\tau=L/Z_0$ is the inductive time constant of the bias tee. We find that nanowire response is well modeled by an expanding hot spot limited by the bias tee supplying the drive current. 

The rise time measured from the homodyne response, defined as the time between reaching $10\%$ and $90\%$ of the full amplitude is shown in Fig.~\ref{fig:homodyne}b. We solve Eqs.(\ref{eqn:resistance_2},\ref{eqn:inductor}) as a function of drive power and fit the results to the data allowing the heat capacity $c$ and thermal conductivity $\alpha$ in Eq.(\ref{eqn:steakly}) to be determined by the fit. We fix the hot spot temperature $T_h=12~\rm{K}$ as the maximum temperature of the hot spot determined in Ref.~\onlinecite{Yang:2007cu}. The model shows good agreement with the data and produces the fitted values $c=1.96\pm0.15\times 10^4~\rm{J/Km^3}$ and $\alpha = 6.20\pm0.43\times 10^5~\rm{W/Km^2}$. The fitted value of the heat capacity is very close to the value predicted by Ref.~\onlinecite{Yang:2007cu} at $12~\rm{K}$ while the value for the thermal contact conductivity is slightly lower due to the fact that this device is fabricated on a thermal oxide substrate rather than sapphire.

The power dissipated by the nanowire can be estimated by noting that the electro-thermal feedback of the hot spot regulates the current in the nanowire to $I_s$, so that the resistance of the nanowire, $R$ is given by $V/I_s-2Z_0$ for an excitation voltage $V$. This relationship implies that to achieve a resistance of $1~\rm{k}\Omega$ the switch must be driven with $\approx200~\rm{nW}$ and will consume $\approx 15~\rm{nW}$ of power. We also note that by increasing the drive voltage and decreasing the critical current, the power consumed by the nanowire can be held constant while the resistance and hence isolation is increased. This is most easily achieved by fabricating long narrow nanowires, but comes at the expense of a decreased saturation power. 

In summary we demonstrate the performance and power consumption of single pole single throw switch based on a superconducting nanowire. In the simplest realization the switch consists of a single nanowire integrated into a superconducting transmission line. Such a switch has a rise and fall time of $\approx 370~\rm{ps}$, consumes $\approx 15~\rm{nW}$ and requires a drive of $\approx 200~\rm{nW}$. The switch exhibits isolation greater than 10 dB from 2 to 8 GHz, limited by the self resonance of the integrated bias inductors. Higher isolations and lower drive powers can be achieved by further miniaturizing and elongating the nanowire switching element. Furthermore it is possible to extend the principle of operation of the tandem nanowire switch design in order to create a single pole double throw switch~\cite{PhysRevApplied.6.024009}. This is achieved when the grounded nanowire in Fig.~\ref{fig:circuit} is replaced by a nanowire connected to a third capacitively coupled port and inductive drive terminal. The switch can be fabricated in a single layer of superconducting material and the wide band, high isolation and low operating power of the nanowire make it well suited to a range of superconducting microwave applications. We anticipate that more complicated and optimized switch designs can be demonstrated in the near future. 

\begin{acknowledgments}
Fabrication of the devices presented in this paper was conducted at the Harvard Center for Nanoscale Systems, a member of the National Nanotechnology Coordinated Infrastructure Network (NNCI), which is supported by the National Science Foundation under NSF award no. 1541959, and Raytheon Integrated Defense Systems at Andover, MA. The research is supported by the Office of the Director of National Intelligence (ODNI), Intelligence Advanced Research Projects Activity (IARPA), under the Cryogenic Computing Complexity program via contract W911NF-14-C0089 as well as internal funding from Raytheon BBN Technologies. The views and conclusions contained herein are those of the authors and should not be interpreted as necessarily representing the official policies or endorsements, either expressed or implied, of the ODNI, IARPA, or the U.S. Government. This document does not contain technology or technical data controlled under either the ITAR or the EAR.
\end{acknowledgments}



%
%

%



\bibliography{rfswitch}
\end{document}